\documentclass[twocolumn,aps,prb,showpacs,preprintnumbers,amsmath,amssymb]{revtex4}
\usepackage{graphicx}  % needed for figures
\usepackage{dcolumn}   % needed for some tables
\usepackage{bm}        % for math
\usepackage{amssymb}   % for math
\usepackage[dvipsnames,usenames]{color}
\usepackage{color}
\usepackage{comment}
\usepackage{gensymb}
\usepackage{physics}

\begin{document}
\newcommand{\etal}{{\sl et al.}}
\newcommand{\ie}{{\sl i.e.}}
\newcommand{\lto}{LaTiO$_3$}
\newcommand{\lao}{LaAlO$_3$}
\newcommand{\lno}{LaNiO$_3$}
\newcommand{\nith}{Ni$^{3+}$}
\newcommand{\nitw}{Ni$^{2+}$}
\newcommand{\otw}{O$^{2-}$}
\newcommand{\alo}{Al$_2$O$_3$}
\newcommand{\aalo}{$\alpha$-Al$_2$O$_3$}
\newcommand{\xto}{$X_2$O$_3$}
\newcommand{\eg}{$e_{g}$}
\newcommand{\tg}{$t_{2g}$}
\newcommand{\dzt}{$d_{z^2}$}
\newcommand{\dxtyt}{$d_{x^2-y^2}$}
\newcommand{\dxy}{$d_{xy}$}
\newcommand{\dxz}{$d_{xz}$}
\newcommand{\dyz}{$d_{yz}$}
\newcommand{\egp}{$e_{g}'$}
\newcommand{\ag}{$a_{1g}$}
\newcommand{\mub}{$\mu_{\rm B}$}
\newcommand{\ef}{$E_{\rm F}$}
\newcommand{\alalo}{$a_{\rm Al_2O_3}$}
\newcommand{\asto}{$a_{\rm STO}$}
\newcommand{\nst}{$N_{\rm STO}$}
\newcommand{\lnnlam}{(LNO)$_N$/(LAO)$_M$}
\newcommand{\lxolao}{(La$X$O$_3$)$_2$/(LaAlO$_3$)$_4$}
\newcommand{\xoalo}{($X_2$O$_3$)$_1$/(Al$_2$O$_3$)$_5$}
\title{Confinement-driven electronic and topological phases  in corundum-derived $3d$-oxide honeycomb lattices}

\author{Okan K\"oksal}
\affiliation{Department of Physics and Center for Nanointegration Duisburg-Essen (CENIDE), University of Duisburg-Essen, Lotharstr. 1, 47057 Duisburg, Germany}
\author{Santu Baidya}
\affiliation{Department of Physics and Center for Nanointegration Duisburg-Essen (CENIDE), University of Duisburg-Essen, Lotharstr. 1, 47057 Duisburg, Germany}
\author{Rossitza Pentcheva}
\email{Rossitza.Pentcheva@uni-due.de}
\affiliation{Department of Physics and Center for Nanointegration Duisburg-Essen (CENIDE), University of Duisburg-Essen, Lotharstr. 1, 47057 Duisburg, Germany}
%\pacs{}
\date{\today}

\begin{abstract}
Using density functional theory calculations including an on-site Coulomb term,
we explore electronic and possibly topologically nontrivial phases in
$3d$ transition metal oxide honeycomb layers confined in the corundum
structure ($\alpha$-Al$_2$O$_3$) along the [0001] direction. 
%Except for $X$?=V, Cr
%and Fe, most of the systems exhibit a ground state that is distinct from
%the corresponding bulk $X_2$O$_3$ compound. In particular, ferromagnetic
%$X$=Ti, Mn, Ni and metastable Co exhibit a characteristic set of four
%bands with a Dirac crossing close to the Fermi level. 
In most cases the ground state is a trivial antiferromagnetic Mott insulator, often with distinct orbital or spin states compared to the bulk phases. With imposed symmetry of the two sublattices the ferromagnetic phases of Ti, Mn, Co and Ni exhibit a  characteristic set of four bands, two relatively flat and two with a Dirac crossing at K, associated with the single electron occupation of $e_{g}'$ (Ti) or $e_{g}$ (Mn, Co, Ni) orbitals. Our results indicate that the Dirac point can be tuned to the Fermi level using strain.
Applying spin-orbit coupling (SOC) leads to a substantial anomalous Hall conductivity
with values up to 0.94 $e^2/h$. Moreover, at $a_{Al_2O_3}=4.81$~\AA\ we identify a particularly strong effect of SOC with out-of-plane easy axis for  ($Ti_2$O$_3$)$_1$/(Al$_2$O$_3$)$_5$(0001) which stabilizes dynamically the system. Due  to the unusually high orbital moment of -0.88$\mu_{\rm B}$ that nearly compensates the spin moment of 1.01 $\mu_{\rm B}$, this system emerges as a candidate for the realization of the topological Haldane model of spinless fermions. Parallels to the perovskite analogs (La$X$O$_3$)$_2$/(LaAlO$_3$)$_4$(111) are discussed.
\end{abstract}

\maketitle

\section{Introduction}
%Confinement is a powerful tool to induce novel behavior that is not available in the bulk. 
Progress in growth techniques like molecular beam epitaxy and pulsed laser deposition has enabled the growth of transition metal oxide (TMO) superlattices with atomic precision. This has opened possibilities to explore emergent phenomena and electronic as well as magnetic phases in reduced dimensions that are not available in the parent compounds \cite{Hwang:12, Mannhart:10, Chakhalian:12,Lorenz:16}. Beyond the [001] growth direction, where confinement can lead to a metal-to-insulator transition e.g. in LaNiO$_3$/LaAlO$_3$(001) superlattices (SLs)\cite{Boris2011,Freeland2011,BlancaRomero2011}, in (111)-oriented  perovskite superlattices  a  buckled honeycomb lattice is formed by each two $X$-cation triangular lattices of the A$X$O$_3$ perovskite structure, as suggested by Xiao et al. \cite{Xiao2011}. Already in 1988 Haldane\cite{Haldane} predicted a quantized Hall conductance arising from spinless fermions  on a honeycomb lattice in the absence of an external magnetic field. This model serves as a prototype of the quantum Hall anomalous insulators (QAHI) that are  of interest for future applications in low-power electronics devices or in the search of Majorana fermions \cite{Weng2015,Ren2016}. Materials realizations of QAHI are being sought in Mn doped HgTe or Cr-, Fe-doped  Bi$_{2}$Te$_{3}$, Bi$_{2}$Se$_{3}$, Sb$_{2}$Te$_{3}$\cite{Liu_Zhang,Yu_Zhang,Fang_Bernevig}, $5d$ transition metals on graphene\cite{Zhang2012,Zhou_Liu} or in TMO with rocksalt- (EuO/CdO\cite{Zhang2014} and EuO/GdO\cite{Garrity2014}) or rutile-derived heterostructures \cite{Huang_Vanderbilt,Cai_Gong,Lado2016} or double perovskites\cite{Cook_Paramekanti,Baidya16}. While the relevant orbitals in common topological insulators are $s$ and $p$, the correlated nature of the $d$ electrons in transition metal oxides suggests a richer functional behavior.

In this context, the initial proposal of Xiao et al.\cite{Xiao2011}  offered a fertile playground to explore the realization of  topologically nontrivial phases in perovskite superlattices with a honeycomb pattern \cite{Ruegg2011,Yang2011,Ruegg2012,Middey2012,Lado2013,Okamoto2013,Doennig2013,Okamoto2014,Doennig2014,Middey2016}. A recent systematic DFT+$U$ study of the $3d$ series in \lxolao(111) superlattices \cite{Doennig2016} revealed a broad set of competing charge, magnetic and orbitally ordered phases. Among the $3d$ series, the LaMnO$_3$ buckled honeycomb bilayer represents a promising candidate with a topological transition from a high-symmetry Chern insulator with a substantial gap (150 meV) to a Jahn-Teller distorted trivial Mott insulating ground state. The insight obtained in this study especially on the effect of band filling and strain has served to identify robust Chern insulators in the $4d$ and $5d$ series (e.g.  LaOsO$_3$ and LaRuO$_3$ bilayers) \cite{HongliNQM}. 
\begin{figure}[h!]
\centering
\includegraphics[width=8.8cm,keepaspectratio]{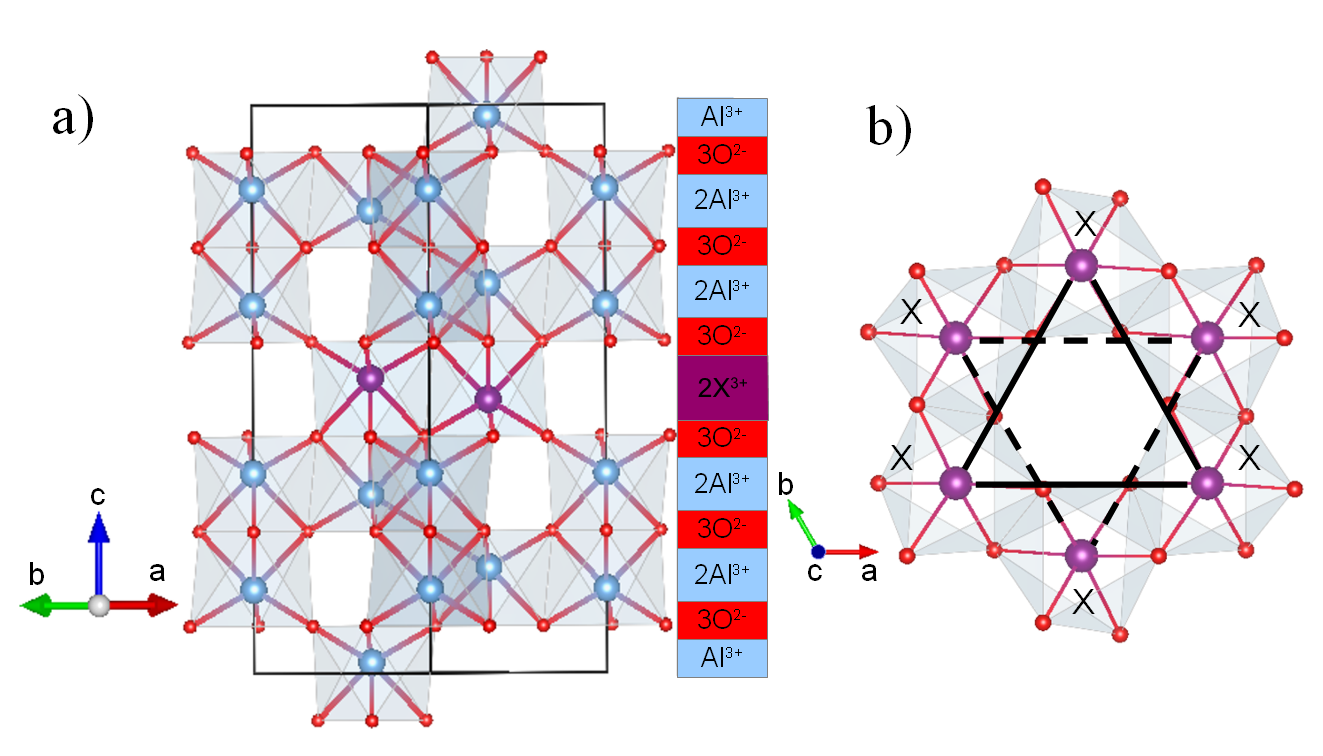}
\caption{(a) Side view of the ($X$$_{2}$O$_{3}$)$_{1}$/(Al$_{2}$O$_{3}$)$_{5}$(0001) superlattice where one metal ion bilayer in the corundum structure is exchanged by a $X=3d$ ion. (b) Top view of the buckled honeycomb lattice in $a$-$b$ plane solid and dashed lines connect the next nearest TM-ion neighbors residing on the two sublattices.} 
\label{fig:schematic_view}
%(c) Edge shared octahedra MO$_{6}$ trigonally distorted. (d) Brillouin zone of hexagonal lattice.
\end{figure}

Besides the (111) oriented perovskite SLs, a honeycomb lattice can be found also in other structure types as e.g. the corundum structure, as mentioned in \cite{Xiao2011}, where each metal ion bilayer forms a honeycomb pattern with a much lower degree of buckling than in the perovskite case. Another qualitative difference is the type of connectivity in the structure. While in the perovskite only corner connectivity is present, in the corundum  the $X$O$_6$ octahedra are edge-sharing in-plane and alternating corner- and face-sharing out-of-plane (cf. Fig. \ref{fig:schematic_view}). This difference in connectivity implies different interaction mechanisms between neighboring sites with potential impact on the electronic and magnetic behavior compared to the perovskite superlattices. Therefore, we perform here a systematic study of the ground state properties of $3d$ honeycomb layers incorporated in the band insulator \aalo, as shown in Fig.~\ref{fig:schematic_view} and compare those to the perovskite analogs\cite{Doennig2016}.  Recently, Afonso and Pardo \cite{Pardo2015} addressed the possibility of topological phases in $5d$ honeycomb layers sandwiched in the corundum structure, concentrating primarily on cases that preserve time reversal and inversion symmetry. In contrast, in the $3d$ cases the electronic correlations are much more pronounced, leading to a stronger affinity towards magnetically ordered states. Here  we take into account the possibility of time-reversal symmetry breaking, resulting in a variety of competing magnetic and electronic phases.  Analogous to the [111]-oriented perovskite heterostructures \cite{Doennig2016}, there is no polar mismatch at the interface between $X_{2}$O$_{3}$ and Al$_{2}$O$_{3}$ and both the $X$ and Al ions are formally in the $3+$ oxidation state (except for the disproportionation found in the case of $X=$ Ni that will be discussed below). We compare the properties of the SLs also to the bulk parent compounds (see more information in Suppl. Mat. ~\cite{suppl}).   %polymorph

The remainder of this paper is structured as follows: In Section \ref{sec_calc} we give a brief description of the computational details and studied systems,  in Section \ref{sec:electronic} we discuss the ground and metastable states within the $3d$ series. In Section \ref{sec:strain} we investigate how the electronic properties can be tuned by strain. In Section~\ref{sec:qahi}  we analyze the effect of spin-orbit coupling (SOC)  and   the resulting anomalous Hall conductivity and Berry curvatures. Moreover, we identify two cases of unusually strong spin-orbit effect. Finally, the results are summarized in Section~\ref{sec:fin}.

\section{Theoretical method and details}\label{sec_calc}
Density functional calculations were carried out for ($X$$_{2}$O$_{3}$)$_{1}$/(Al$_{2}$O$_{3}$)$_{5}$(0001), $X= 3d$ ion, with the VASP\cite{VASP} code using pseudopotentials and  projector augmented waves (PAW) \cite{PAW}. For the exchange-correlation functional we applied the generalized gradient approximation (GGA) of  Perdew, Burke and Enzerhof \cite{GGA_PBE} and in some cases for comparison the local density approximation (LDA)~\cite{suppl}. An on-site Coulomb repulsion parameter was considered within the GGA+$U$ approach with $U$\,=\,5 eV and Hund's exchange parameter of \textit{J}\,=\,0.7 eV on all TM ions $X$. We used the approach of Dudarev et al.\cite{Dudarev}, which considers $U_{\rm eff}=U-J$. We have verified that the results are robust   w.r.t. a reasonable variation of $U$, and that LDA+$U$ and GGA+$U$ give qualitatively similar behavior, further information is  provided in Suppl. Material~\cite{suppl}. Previous studies have shown that taking into account electronic correlations may result in a trivial AFM insulator for $5d$ systems, initially proposed as candidates for a $Z_2$ TIs within tight binding and DFT (GGA/LDA) studies. While for a SrIrO$_3$ honeycomb bilayer both DFT+DMFT \cite{Okamoto2014} and DFT+$U$ \cite{Lado2013} show such a tendency, the effect of static vs. dynamic correlations needs to be addressed in more detail in future studies. 

The $k$-point mesh contains at least 6$\times$6$\times$2 $k$-points including the $\Gamma$ point, in some cases 9$\times$9$\times$3 $k$-points were used. A cutoff energy of 600 eV was chosen for the plane waves. Atomic positions were relaxed  until  the Hellman-Feynman forces were lower than 1 meV/\AA.  In selected cases, especially in order to explore high-symmetry metastable states or symmetry-broken states, calculations were performed using the all-electron full-potential linearized augmented plane wave (LAPW) method as implemented in the Wien2k code\cite{wien2k}. 
The anomalous Hall conductivity (AHC) for potentially interesting cases is computed on a very dense $k$-point mesh of 300$\times$300$\times$50 using the wannier90 code\cite{wannier90}. %For the construction of the maximally localized Wannier functions\cite{Marzari_Vanderbilt} the \tg\ and \eg\ manifold of the $3d$ TM cations as well as the O-$2p$ orbitals  are considered.

%\section{Crystal Structure}
To model the ($X$$_{2}$O$_{3}$)$_{1}$/(Al$_{2}$O$_{3}$)$_{5}$(0001) superlattices, an Al-bilayer in the corundum structure is substituted by a $X=3d$ bilayer (cf. Fig. \ref{fig:schematic_view}). Systems grown on \aalo(0001) are simulated by fixing the lateral lattice constant to the GGA value of \alo. The lattice parameters of \aalo\ within GGA are $a$=\,4.81\,\AA,\,$c$\,=\,13.12 \AA, approx. 1\%  larger  than the experimental values $a$=\,4.76\,\AA\,and \,$c$\,=\,12.99 \AA \cite{Newnham1962}. The internal parameters and the out-of-plane lattice parameter $c$ were optimized for each system within GGA+$U$ and the results  are given in Table II in  Suppl. Material \cite{suppl}. Additionally, we calculated the structural and electronic properties of bulk $X_2$O$_3$, displayed in Tab. I in Suppl. Material\cite{suppl}. The lateral strain is defined as $\sigma(\%)=\frac{\Delta a}{a_{X_{2}O_{3}}}$ where $\Delta a = a_{Al_{2}O_{3}}-a_{X_{2}O_{3}}$ is the difference between the lateral lattice parameters of \alo\ and bulk $X_2$O$_3$. We have considered both ferro- (FM) and antiferromagnetic (AFM) coupling and performed the relaxations for each spin arrangement for both the bulk compounds and the SLs. As mentioned above, the degree of buckling is much lower for the corundum-derived honeycomb layers (0.39-0.63~\AA, see Table II in Suppl. Material\cite{suppl}) compared to the perovskite analogs (2.27-2.44~\AA)\cite{Doennig2016}.

\section{Electronic and magnetic properties}
\label{sec:electronic}
\begin{figure*} [htb!]
\includegraphics[width=12cm,keepaspectratio]{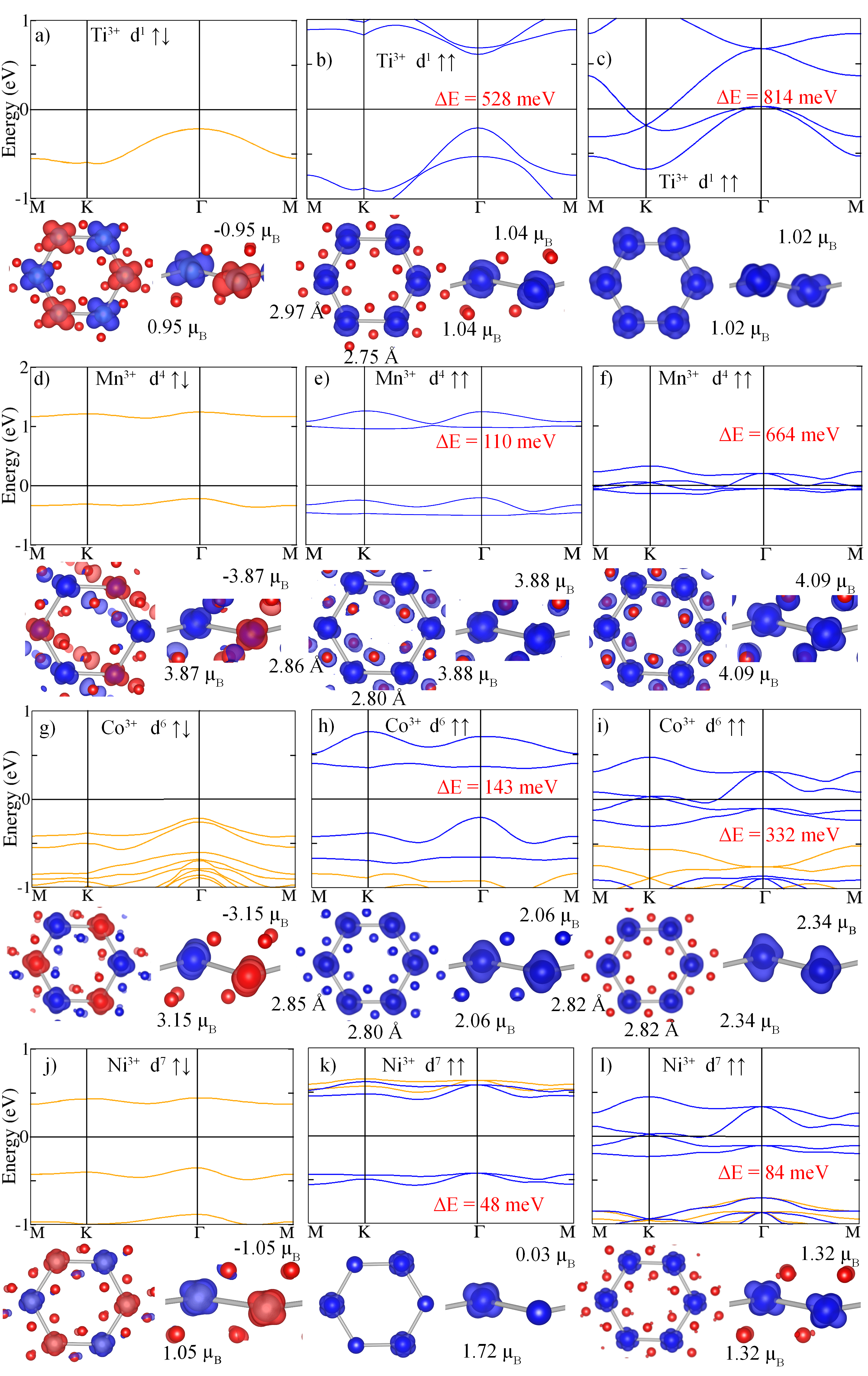}
\caption{Band structures and spin density distributions of \xoalo(0001) with $X$\,=\,Ti,\,Mn,\,Co,\,Ni for FM and AFM coupling. The spin density is integrated in the energy range between --8 eV to \ef, except for d-f) and j) where the integration interval is  --0.6 eV to \ef.  In the band structure blue/orange denote majority/minority bands and the Fermi level is set to zero. In the isosurfaces of the spin density blue (red) show the  majority (minority) contributions. The energies of metastable states are given in red, also the magnitude of the spin moments of the $X$ ions are displayed in the spin-densities.}
\label{fig:t2g_eg_systems}
\end{figure*}

\begin{figure} [h!]
\centering
\includegraphics[width=8.5cm,keepaspectratio]{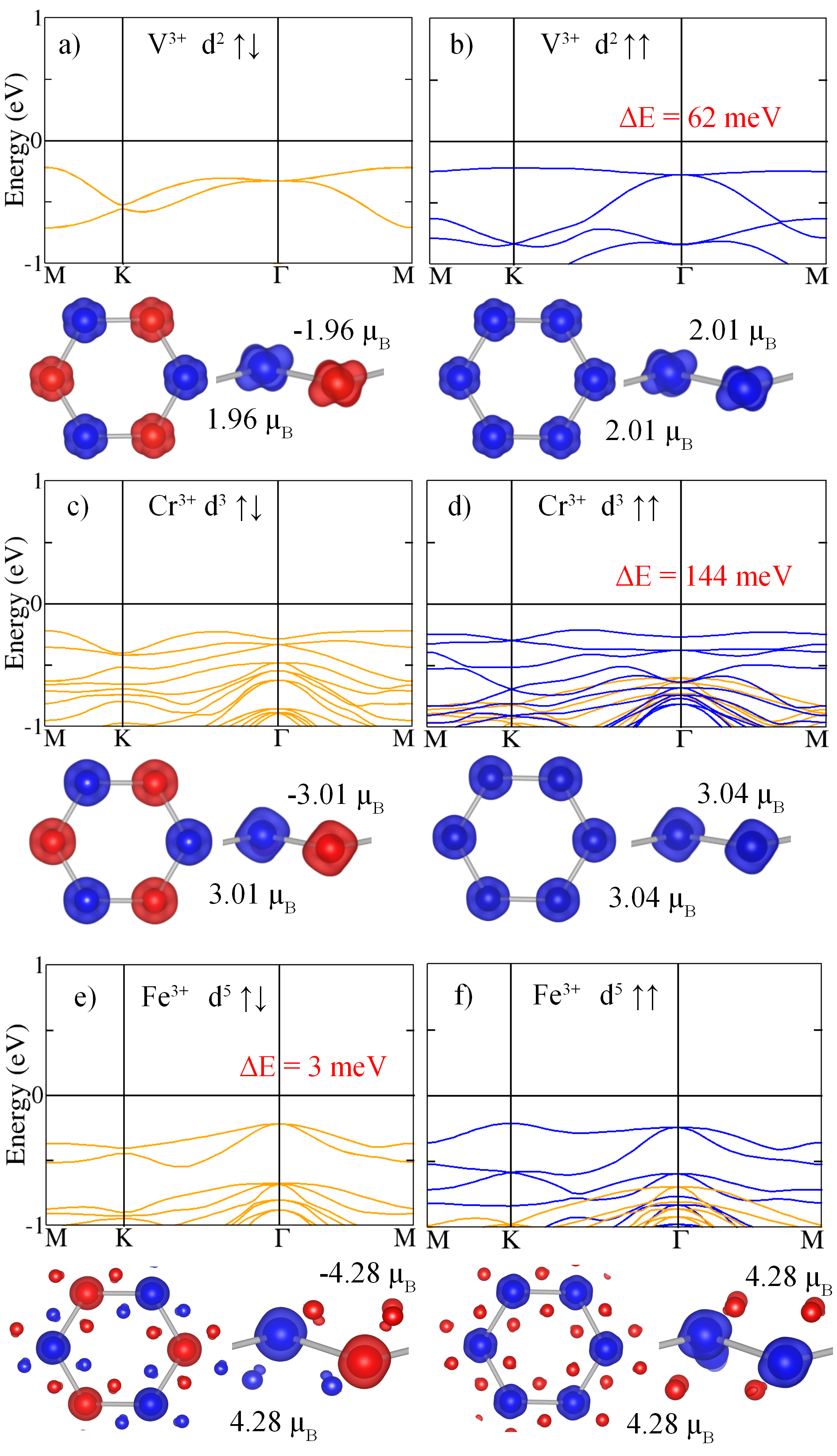}
\caption{Band structures and spin density distributions integrated in the energy range between  --8 eV to \ef\ for $X$\,=\,V,\,Cr, and between --0.6 eV to \ef\ for Fe with FM and AFM coupling. Same colour coding is used as in Fig.~\ref{fig:t2g_eg_systems}.}
\label{fig:t2g_systemsVCrFe}
\end{figure}

In this Section we discuss the electronic and magnetic properties of stable and metastable ($X_{2}$O$_{3}$)$_{1}$/(Al$_{2}$O$_{3}$)$_{5}$(0001) superlattices with AFM and FM order.  Among the $3d$ series we have identified four cases $X$\,=\,Ti,\,Mn,\,Co,\,Ni with common features whose band structures and spin-densities  are shown in Fig. \ref{fig:t2g_eg_systems}. The remaining systems are discussed towards the end of this Section (cf. Fig. \ref{fig:t2g_systemsVCrFe}). The four above mentioned cases exhibit an AFM insulating ground state. In the ferromagnetic cases with constrained symmetry of the two sublattices, we identify a common band structure with a characteristic set of four bands around \ef. Two of these bands are flat and nearly dispersionless, while the other two show a Dirac-like crossing at K close to \ef. This bears similarities to the perovskite analoga \lxolao(111) with $X$\,=\,Mn,\,Co,\,Ni and originates from the single occupation of the \egp\ and \eg\ manifold, respectively, as will be discussed below. Distinct to the perovskite case, however, the Dirac crossing is not exactly at \ef, but $\sim200$ meV below (Ti) or slightly above (Mn, Co and Ni) the Fermi level. The resulting electron (hole) pockets are compensated by hole (electron) pockets around $\Gamma$ (Ti) or along K-$\Gamma$ and $\Gamma$-M for the remaining cases. In Section  \ref{sec:strain} we will explore whether these coupled electron-hole pockets can be quenched and the Dirac point shifted to the  Fermi level using epitaxial strain. A further notable feature is the variation of bandwidth of those bands: while for Ti the bandwidth is substantial ($\sim2$ eV, cf. Fig. \ref{fig:t2g_eg_systems}c), for $X$\,=\,Mn,\,Co,\,Ni it is much narrower ($0.5-0.8$ eV, cf. Fig. \ref{fig:t2g_eg_systems}f, i, l). In all four FM cases symmetry breaking stabilizes insulating states with significant band gaps (Fig. \ref{fig:t2g_eg_systems}b, e, h, k). Moreover, both for FM and AFM order we observe a rich variety of orbital reconstructions visible from the spin density of the relevant bands that are distinct from the bulk, which we discuss in more detail in the following. 

In the corundum structure the symmetry of the $X$O$_{6}$ octahedron is reduced from  octahedral to trigonal which splits the \tg-orbitals into \ag\ and \egp. Similar to the perovskite analog LaTiO$_3$\cite{Doennig2016}, the Ti$_2$O$_3$ honeycomb layer reveals a competition between the cubic \tg-orbitals (\dxy, \dxz, \dyz) and the above mentioned trigonal ones. In particular, a \dxy-shaped orbital is stabilized in the AFM ground state (Fig. \ref{fig:t2g_eg_systems}a), while the symmetry-broken FM state exhibits a linear combination of two orbitals (e.g. \dyz+\dxz, cf. Fig. \ref{fig:t2g_eg_systems}b)). On the other hand, the FM case with constrained symmetry of the sublattices has an \egp\ orbital polarization (Fig. \ref{fig:t2g_eg_systems} c). For comparison, in the perovskite LaTiO$_3$-bilayer the FM ground state comprises a staggered order of cubic orbitals (\dxz, \dyz)\cite{Doennig2016}, whereas in the corundum case the same orbital polarization occurs on both sublattices, likely related to the different (edge vs. corner) connectivity in the corundum-derived honeycomb layer. It is noteworthy, that these orbital polarizations are at variance with the \ag\ occupation in  bulk Ti$_2$O$_3$\cite{suppl}.  As a result of the  symmetry breaking and orbital order a gap opens for the AFM ground state (2.33 eV) and the FM state (0.81 eV) which lies 528 meV higher in energy. In contrast, the FM state with symmetric sublattices is 814 meV less stable owing to the different orbital polarization (single electron in the doubly degenerate \egp\ orbitals)  and the resulting metallic phase with the above described set of four majority-spin bands.  It would be interesting to explore further these states e.g. with DFT+ DMFT.  
%, two of which are crossing at the K point (cf. Fig. \ref{fig:t2g_eg_systems} c). As we will see in the following, this feature occurs for several of the discussed corundum SLs originating from the single occupation of the \egp\ and \eg\ manifold, respectively. We note that the Dirac-like crossing at K lies $\sim200$ meV below the Fermi level.  This is compensated by hole pockets at $\Gamma$. In Section  \ref{sec:strain} we will explore if this coupled electron-hole pocket can be quenched and the Dirac point shifted to the  Fermi level and how the orbital polarization can be tuned by strain.

The AFM ground state of $X=$ Mn ($d^4$) is insulating with a gap of 1.35 eV that separates very flat bands (see Fig. \ref{fig:t2g_eg_systems}d). Its origin is  a Jahn-Teller (JT) distortion which lifts the degeneracy of the singly occupied \eg\ band and reduces the symmetry to P1. Due to the superposition of trigonal and JT distortion the relaxation pattern is more complex, the Mn-O distances change from  1.98, 2.10 \AA\ (symmetric FM case) to apical distances of 2.21 and 2.09 \AA\ and basal ones varying between 1.92 and 2.03 \AA\ for the JT distorted case.     The spin-density integrated between --0.6 eV to \ef\ indicates a  \dzt-shaped orbital polarization and a strong hybridization with O$2p$ states. We note that the orbitals on both sublattices have the same orientation, in contrast to the perovskite JT-distorted (LaMnO$_3$)$_2$/(LaAlO$_3$)$_4$(111) case, which shows alternating orientation of the \dzt-shaped orbitals on the two sublattices \cite{Doennig2016}.  Starting from the FM solution with symmetric sublattices (Fig. \ref{fig:t2g_eg_systems}f), which is 0.66 eV/u.c. higher in energy, a similar orbital polarization as in the AFM case  arises that opens a gap of 1.20 eV in the initially metallic set of four bands (Fig. \ref{fig:t2g_eg_systems}e). This case  lies 110 meV higher in energy than the AFM ground state. 
%For FM coupling with for Ti emerges around the Fermi level. In contrast to the LaMnO$_3$(111) bilayer, where the Dirac-point is fixed at \ef\ in the symmetric case\cite{Doennig2016}, in (Mn$_{2}$O$_{3}$)$_{1}$/(Al$_{2}$O$_{3}$)$_{5}$(0001) the Dirac-crossing at K is above \ef, creating a hole pocket, which is compensated by electron pockets along K-$\Gamma$ and $\Gamma$-M.  Again in Section \ref{sec:strain} we will address, how these electron-hole pockets can be suppressed by strain.

$X=$  Co and Ni  show also a number of intriguing stable and metastable states. While bulk Co$_2$O$_3$ is an insulator with Co $^{3+}$ ($d^6$) in a low spin state ($S=0$) ~\cite{suppl},  the ground state of the confined honeycomb corundum-derived lattice is  an antiferromagnetic insulator with a  band gap of 1.61 eV  and Co in the high spin state (calculated magnetic moment of 3.15 \mub, Fig. \ref{fig:t2g_eg_systems}g) with a full $d$ band in one spin direction and a single electron in the other.   For ferromagnetic coupling two metastable states arise: With imposed symmetry of the two sublattices in  the honeycomb layer the previously observed set of four bands emerges (Fig. \ref{fig:t2g_eg_systems}i). This state is 332 meV less stable than the ground state. When the symmetry constraint is lifted, a band gap opens separating the two occupied from the two empty bands, 143 meV above the ground state (cf. Fig. \ref{fig:t2g_eg_systems}h). Remarkably, the symmetry breaking mechanism here is a dimerization of the Co sites, manifested in alternating Co-Co distances of 2.80~\AA\ and 2.85~\AA, in contrast to the 2.82~\AA\ in the symmetric case (note also the alternating $X-X$ distances for $X=$Ti and Mn, Fig.  \ref{fig:t2g_eg_systems}b and e, respectively). 

The ground state of (Ni$_{2}$O$_{3}$)$_{1}$/(Al$_{2}$O$_{3}$)$_{5}$(0001) is also an AFM insulator with a band gap of $\sim 0.69$ eV which separates two nearly flat bands (see Fig. \ref{fig:t2g_eg_systems}j). If for FM coupling the symmetry of the two sublattices is constrained, a metastable state occurs 84 meV above the ground state with the familiar set of four bands and a Ni-magnetic moment of 1.32 \mub\ (cf. Fig. \ref{fig:t2g_eg_systems}l), similar to the perovskite-derived LaNiO$_3$ bilayer\cite{Yang2011,Ruegg2011,Doennig2014,Doennig2014}. Releasing the symmetry constraint results in a ferromagnetic insulating state with a gap of $\sim 1.0$ eV which arises due to disproportionation of the two Ni sublattices, reflected in a large magnetic moment 1.72~\mub\ on one site and a nearly quenched one on the second, 0.03~\mub\ (cf. Fig. \ref{fig:t2g_eg_systems}k). This is accompanied by a bond disproportionation with Ni-O bond lengths of  2.05, 1.97 \AA\ at the first and 1.91, 1.92 \AA\ at the second site. The corresponding NiO$_6$ volumes are 10.6 and 9.2~\AA$^3$, respectively. This behavior bears analogies to the site-disproportionation in rare earth nickelates\cite{millis,sawatzky} and LaNiO$_3$/LaAlO$_3$(001)  and (111) SLs \cite{Boris2011,Freeland2011,BlancaRomero2011,Doennig2014}. %+The antiferromagnetic ground state exhibits . 
%The antiferromagnetic case is on the other hand  metallic and 473 meV above the ground state (Fig. \ref{fig:t2g_eg_systems}d). 

The band structures and spin densities for the remaining $X$\,=\,V,\,Cr,\,Fe with FM and AFM coupling are shown in Fig. \ref{fig:t2g_systemsVCrFe}. For $X=$V ($d^2$), the \egp\ doublet is fully occupied and both the AFM and FM cases are insulating with broad band gaps (e.g. 2.70 eV for AFM). For FM alignment, the  set of four bands that is half-occupied for Ti  is fully occupied and below \ef\ (cf. Fig. \ref{fig:t2g_systemsVCrFe} b). Both V and Cr ($d^3$ with a half-filled \tg\ subset) are trivial AFM insulators (cf. Fig. \ref{fig:t2g_systemsVCrFe} a and c). In both cases the band gaps are significantly increased compared to the bulk cases due to confinement (cf. Table I  and II in Suppl. Material\cite{suppl}).

Interestingly, $X$=Fe ($d^5$) shows almost degenerate FM and AFM states with a slight preference for FM coupling by 3 meV/u.c., indicating weak exchange interaction. This is consistent with bulk $\alpha$-Fe$_2$O$_3$ that has small AFM in-plane magnetic interaction parameters (note that the strongest AFM coupling in bulk $\alpha$-Fe$_2$O$_3$ occurs to the next layer\cite{hasanPRB,hasanPRB2}, which is quenched in the SL  due  to the presence of Al ions). For the FM case of $X=$ Fe the set of four bands is fully occupied and lies 0.2 eV below \ef\ (Fig.~\ref{fig:t2g_systemsVCrFe}f).  

\section{Effect of strain}\label{sec:strain}
\begin{figure} [htb!]
\centering
\includegraphics[width=9.0cm,keepaspectratio]{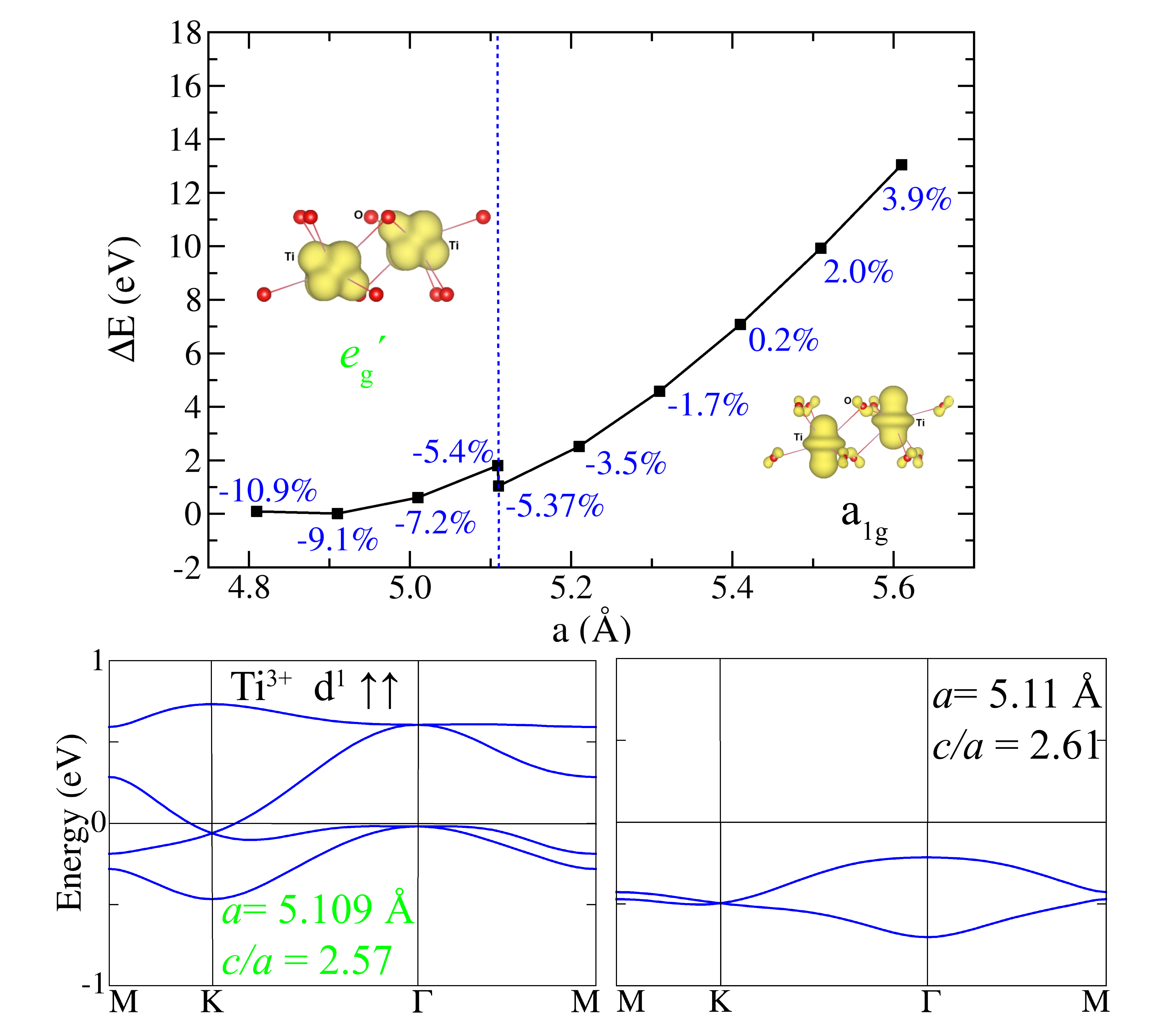}
\caption{Energy difference w.r.t.  the ground state per u.c. vs. in-plane lattice constant for $X$\,=\,Ti. For each value of the lateral strain, the out-of-plane lattice constant was optimized within GGA+$U$. Additionally, the band structure is shown prior and after the switching of orbital polarization from \egp\ to \ag\ at $a= 5.11$\AA.}
\label{fig:strainTi}
\end{figure}
%Additionally, the band structures and the orbital polarization before and after the sharp transition from \egp\ to \ag\ orbital polarization at $a= 5.11$\AA\ are shown. 
%here sensitivity to strain für Ti erwähnen\label{fig:strainTi}

\begin{figure} [ht!]
\centering
\includegraphics[width=9.0cm,keepaspectratio]{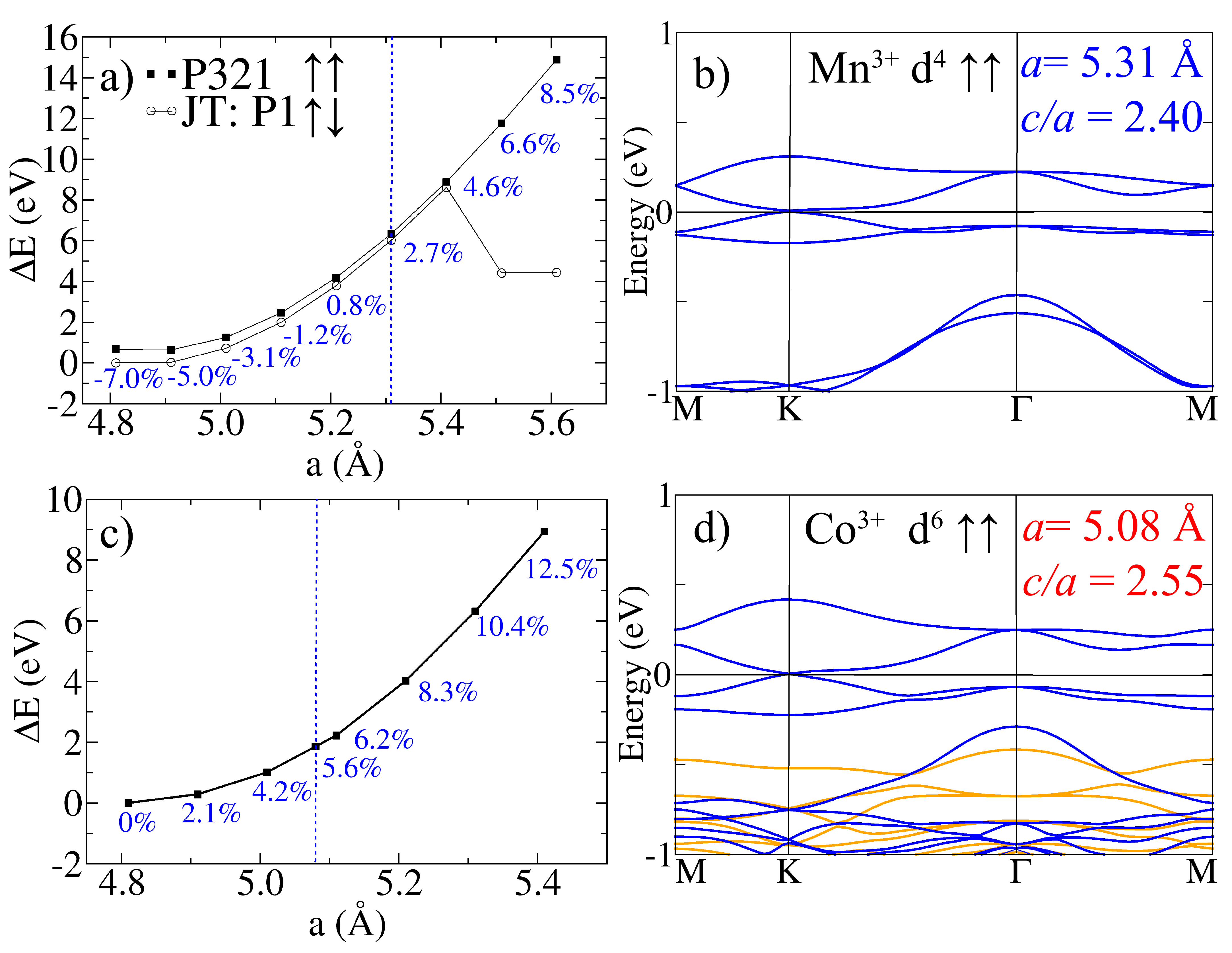}
\caption{Energy difference w.r.t. the ground state per u.c. vs. in-plane lattice constant for  a) $X$\,=\,Mn (closed and open symbols denote FM and AFM coupling) and c) $X$=\,Co. For each value of the lateral strain, the out-of-plane lattice constant was optimized within GGA+$U$. b) and d) show the band structures at the optimum lattice parameters with the DP tuned to \ef.  }
%here sensitivity to strain für Ti erwähnen
\label{fig:strainMnCo}
\end{figure}

As discussed above, in  the symmetric ferromagnetic phases of $X=$ Ti, Mn Co and Ni  a set four bands with a linear crossing at K is observed. However, this Dirac-like crossing is either slightly below (Ti) or slightly above (Mn, Co, Ni) the Fermi level when the system is strained at the lateral lattice constant of \aalo.  In this Section we explore if it is possible to tune the position of the Dirac-point (DP) to the Fermi level using strain. The energy vs. lateral lattice constant   and the band structure at the optimum lattice parameter are plotted in Figs.~\ref{fig:strainTi} and \ref{fig:strainMnCo}. For $X=$Ti the compensation of the electron and hole pockets at K and $\Gamma$ is only partially possible, as at $a=5.11$ \AA\ a sharp transition from \egp\  to \ag\ orbital polarizations takes place.  The band structures and electron density distributions of the bands between -1 eV and \ef\ just prior and after the transition are shown in Fig. \ref{fig:strainTi}. 

The stability of the FM and AFM  phases of  $X=$Mn as a function of strain is displayed in Fig~\ref{fig:strainMnCo}a. While the AFM phase corresponds to the ground state for compressive strain, the energy difference to the FM state is reduced with increasing tensile strain. Moreover, a complete compensation of the electron and hole pockets  can be achieved for the FM state under tensile strain for $a=5.31$~\AA\ (Fig.~\ref{fig:strainMnCo}b).  For the metastable Co-bilayer this state is reached at a  smaller lateral lattice parameter, $a=5.08$~\AA\  (Fig. \ref{fig:strainMnCo}d). Finally, for $X=$ Ni this state occurs for a lateral lattice constant of $a=4.94$~\AA\ just slightly above \alalo (not shown here). Overall, in the latter three cases the DP can be readily tuned to the Fermi level by tensile strain, whereas the level of strain decreases with increasing band-filling. We note that due to the well-known overestimation of lattice constants within GGA+$U$ the actual transitions may occur at smaller lattice constants.

\section{Effect of spin-orbit coupling and anomalous Hall conductivity}
\label{sec:qahi}
%underlying physics
\begin{figure} [ht!]
\centering
\includegraphics[width=8.6cm,keepaspectratio]{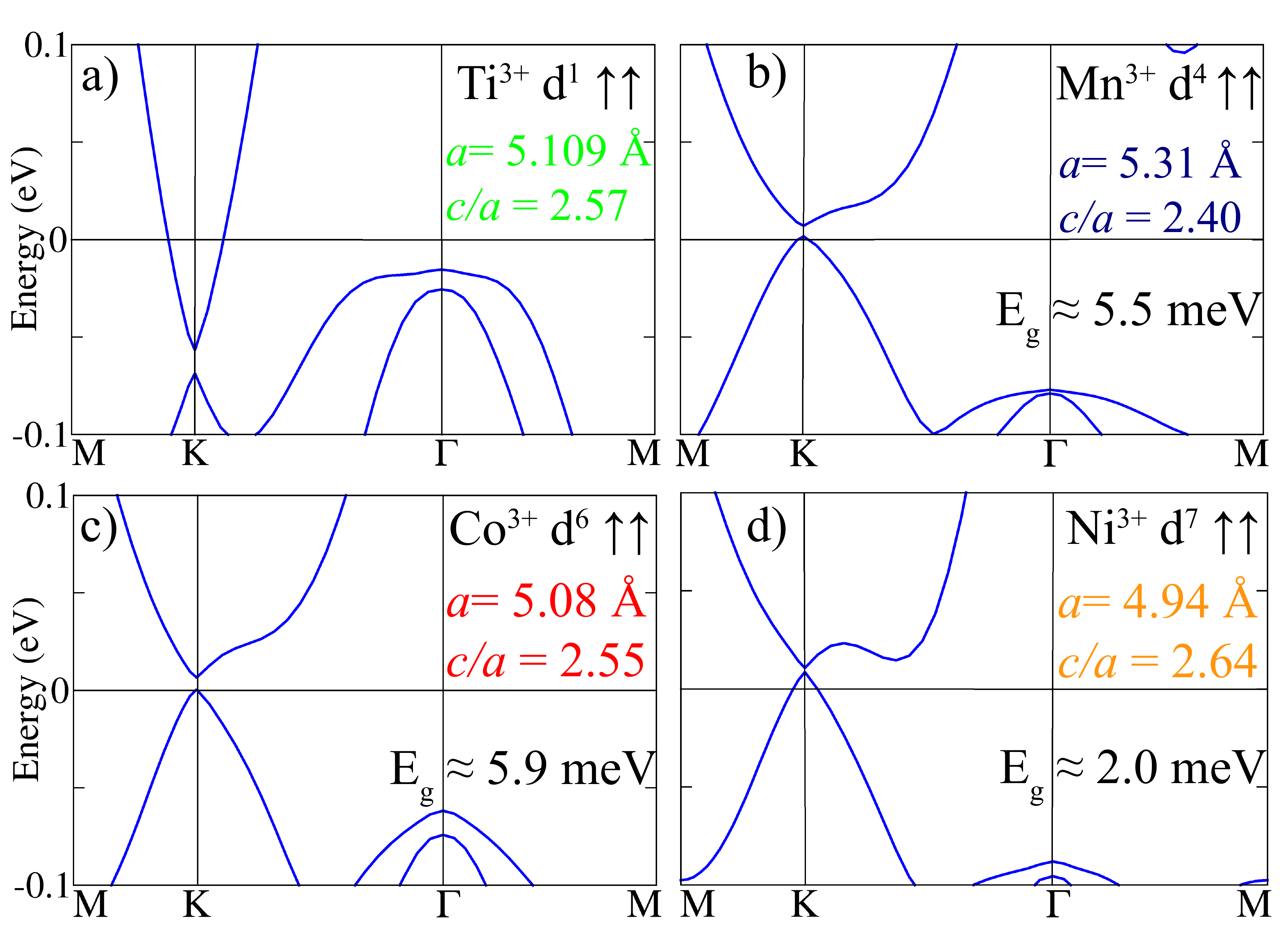}
\caption{GGA\,+\,\textit{U}\,+\,SOC band structures for \xoalo(0001)  with $X$\,=\,Ti,\,Mn,\,Co,\,Ni at the optimum lateral lattice constant and with magnetization  along the [0001] direction. SOC opens a small gap at the DP at K. With the exception of $X$\,=\,Ti and Ni, the \ef\ is found to be located within the gap resulting in a topologically non-trivial gapped state.}
\label{fig:gga_u_soc}
\end{figure}
%The inclusion of SOC in (0001) direction is responsible 
%As shown in the previous section a set of four bands with a Dirac-like crossing at K are obtained for $X$\,=\,Ti, Mn, and metastable Co and Ni. The Dirac point can be tuned to the Fermi level by strain in all above mentioned cases except for Ti, where a transition from \egp\ to \ag\ orbital polarization obstructs the complete shift of the DP to \ef. 
The band structures of the ferromagnetic cases of $X$\,=\,Ti, Mn, Co and Ni with symmetric sublattices indicate a possible topologically nontrivial character which we analyze in the following. In a first step we explore the effect of spin-orbit coupling for the optimum lateral lattice constants obtained in the previous Section and the resulting band structures are displayed in Fig.~\ref{fig:gga_u_soc}.  SOC leads to a gap opening  at the former DP  of a few meV ($\sim$ 2.0-5.9 meV for the Mn and Co corundum bilayers, respectively),  for Ti and Ni the avoided crossing is 50 meV below and  6 meV above \ef, respectively. We note that here the effect of SOC appears  to be much smaller than for the  (111)-oriented \lxolao(111), where SOC opens a gap of $\sim$150 meV for LaMnO$_3$\cite{Doennig2016}. Likewise, the orbital moments are also small and range from 0.01\mub\ (Mn) to 0.03\mub\ (Ti) and 0.04\mub\ (Co and Ni).

For the analysis of the topological properties we project the bands around \ef\ to Wannier functions (WF) using %the wien2wannier interface\cite{wien2wannier} an 
the wannier90 code\cite{wannier90}. To achieve a better localization of the WF, we have chosen a larger energy window for the Wannier interpolation, that includes besides $X3d$ also O$2p$ states. The Berry curvature $\Omega$(k) is calculated through summation over all occupied bands below the Fermi level \cite{Wang_Vanderbilt}: 
\begin{equation} \label{eq:1}
%\resizebox{0.9\hsize}{!} 
{\Omega(k) = -\sum_{n<E_{F}} \sum_{m\neq n} 2Im \frac{ \left\langle \Psi_{nk} \left| v_{x} \right| \Psi_{mk} \right\rangle
\left\langle \Psi_{mk} \left| v_{y} \right| \Psi_{nk} \right\rangle}{{({\epsilon}_{mk}-{\epsilon}_{nk})}^{2}}}				
\end{equation}
where $\Psi_{nk}$ represents the spinor Bloch wave function of band $n$ with an eigenenergy $\epsilon_{nk}$
and $v_x$, $v_y$ are the components of the anomalous velocity.  The anomalous Hall conductivity (AHC) is computed
by integrating the Berry curvature $\Omega(k)$, weighted by an occupation factor $f_{n}(k)$, over the BZ using a dense $k$-point grid of $300\times300\times50$: 
\begin{equation}\label{eq:2}
%\resizebox{0.6\hsize}{!} 
{\sigma_{xy}= -\frac{e^{2}}{2\pi h} \sum_{n} \int_{BZ} dk f_{n}(k)  \Omega_{n,z} (k)}
\end{equation}
%$\sigma_{xy}$\,=\,--$\sigma_{yx}$ defines the anti-symmetric part of the conductivity. \cite{Wang_Vanderbilt,Nagaosa} %The BZ integral of the Berry curvature is weighted by an occupation factor $f$$_{n}$ (k) with respect to each state whereas the Berry curvature is given for $n$th band at K, respectively. 
\begin{figure} [ht!]
\centering
\includegraphics[width=8.2cm,keepaspectratio]{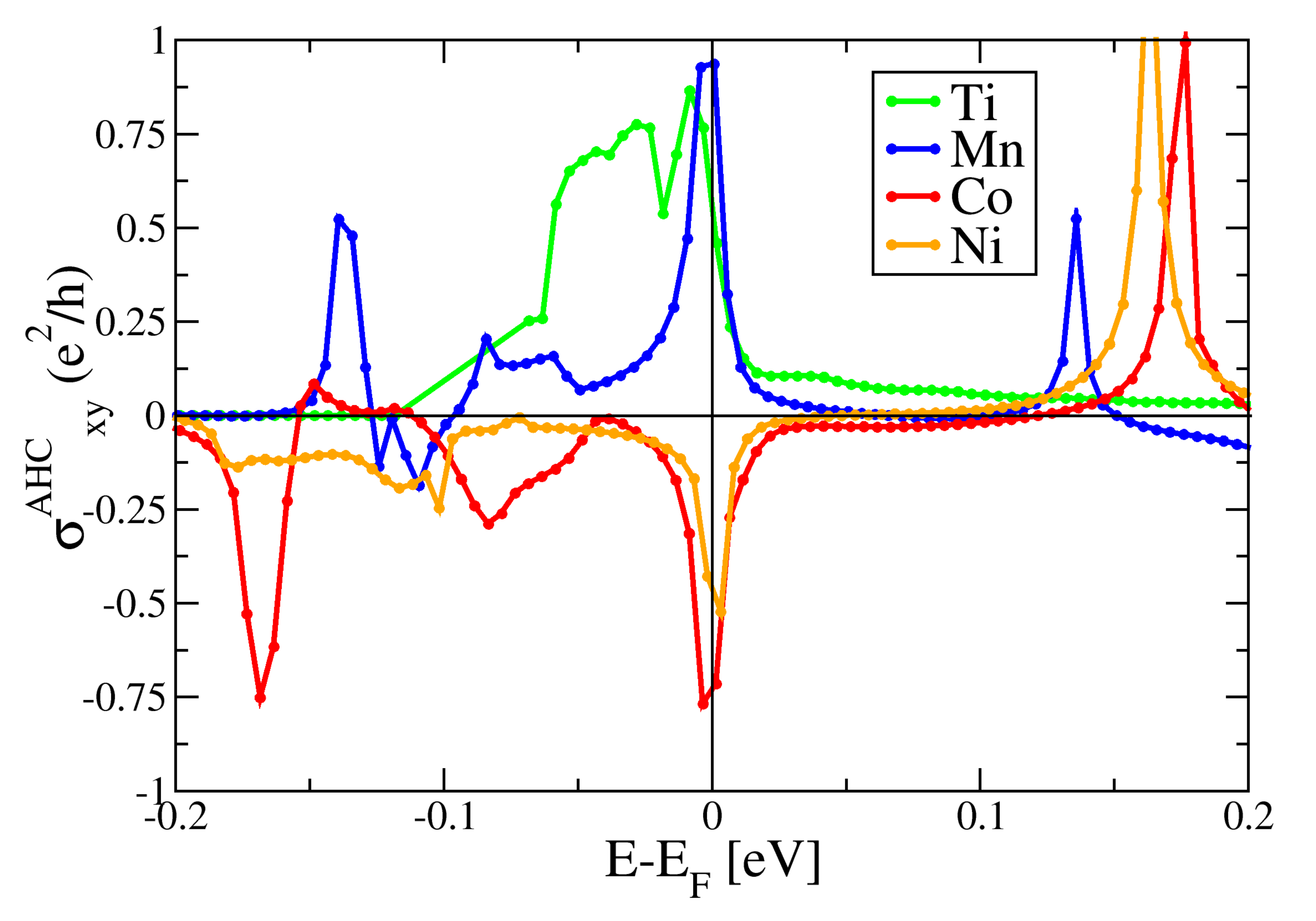}
\caption{AHC $\sigma_{xy}$ in units of $e^{2}/h$ as a function of the chemical potential for the corundum bilayers ($X_{2}$O$_{3}$)$_{1}$/(Al$_{2}$O$_{3}$)$_{5}$(0001) ($X$\,=\,Ti,\,Mn,\,Co,\,Ni).}
%Hence, it can be claimed that $\sigma$$_{xy}$ tends to give a topologically non-trivial state going to an integer value of 1. }
\label{fig:AHC}
\end{figure}
\begin{figure} [ht!]
\includegraphics[height=5cm,keepaspectratio]{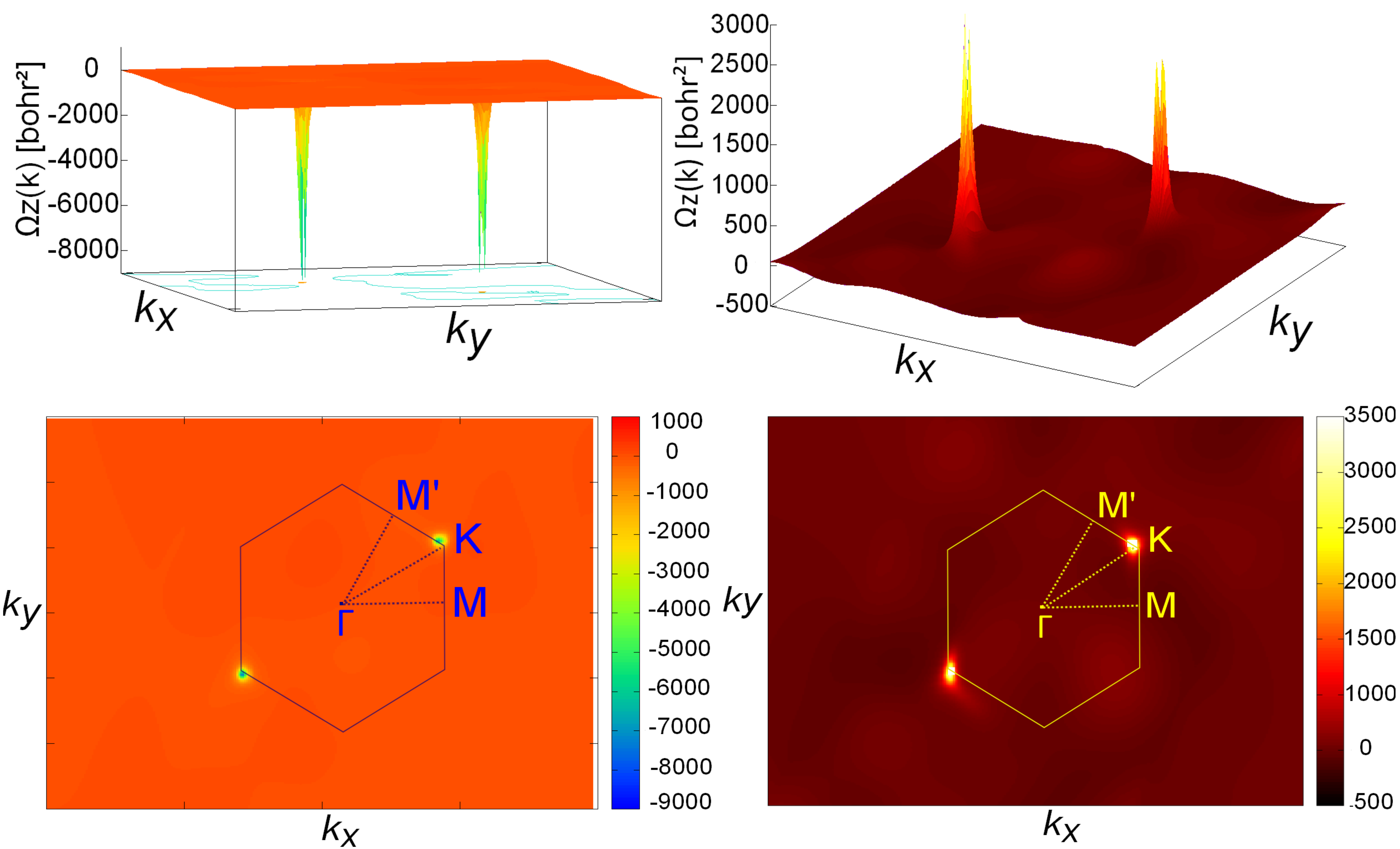}
\caption{Side and top views of the Berry curvature distribution $\Omega_{z}(k_x,k_y)$ for  (Co$_{2}$O$_{3}$)$_{1}$/(Al$_{2}$O$_{3}$)$_{5}$(0001) (left) and (Mn$_{2}$O$_{3}$)$_{1}$/(Al$_{2}$O$_{3}$)$_{5}$(0001) (right). Note the opposite sign of $\Omega_{z}(k_x,k_y)$ for the two cases. 
%The the Brilloihexagon shape outlined by the solid lines in the top views illustrate the first BZ showing the two K and $K'$.
}
\label{fig:Berry_curvature}
\end{figure}
The anomalous Hall conductivity $\sigma_{xy}$ of  the corundum bilayers with the chemical potential  varying $\pm 0.2$ eV around \ef\ is displayed in Fig. \ref{fig:AHC}. All four systems show significant values approaching $+1$ (Ti and Mn) or $-1$ (Co and Ni). In particular, (Mn$_{2}$O$_{3}$)$_{1}$/(Al$_{2}$O$_{3}$)$_{5}$(0001) and (Co$_{2}$O$_{3}$)$_{1}$/(Al$_{2}$O$_{3}$)$_{5}$(0001)  exhibit values of $\sim 0.94$ and $\sim -0.78 e^{2}/h$, respectively.  In contrast to other systems, like e.g.  TiO$_{2}$/VO$_{2}$ \cite{Huang_Vanderbilt}, the plateaus here are very narrow, of the order of the small band gap opened by SOC. Hence a small denominator and large velocity matrix elements in Eq. \ref{eq:1} give rise to  large contributions to the Berry curvature $\Omega(k)$ as shown in Fig. \ref{fig:Berry_curvature} where sharp peaks arise around K with values of 3000 and --8000 bohr$^{2}$  for (Mn$_{2}$O$_{3}$)$_{1}$/(Al$_{2}$O$_{3}$)$_{5}$(0001) and (Co$_{2}$O$_{3}$)$_{1}$/(Al$_{2}$O$_{3}$)$_{5}$(0001), respectively. The different signs in the Berry curvature correlate with the negative and positive signs of the Hall conductivity.

%\section{Strong Effect of Spin-Orbit Coupling}
%\label{sec:soc}
\begin{figure} [ht!]
\centering
\includegraphics[width=7.0cm,keepaspectratio]{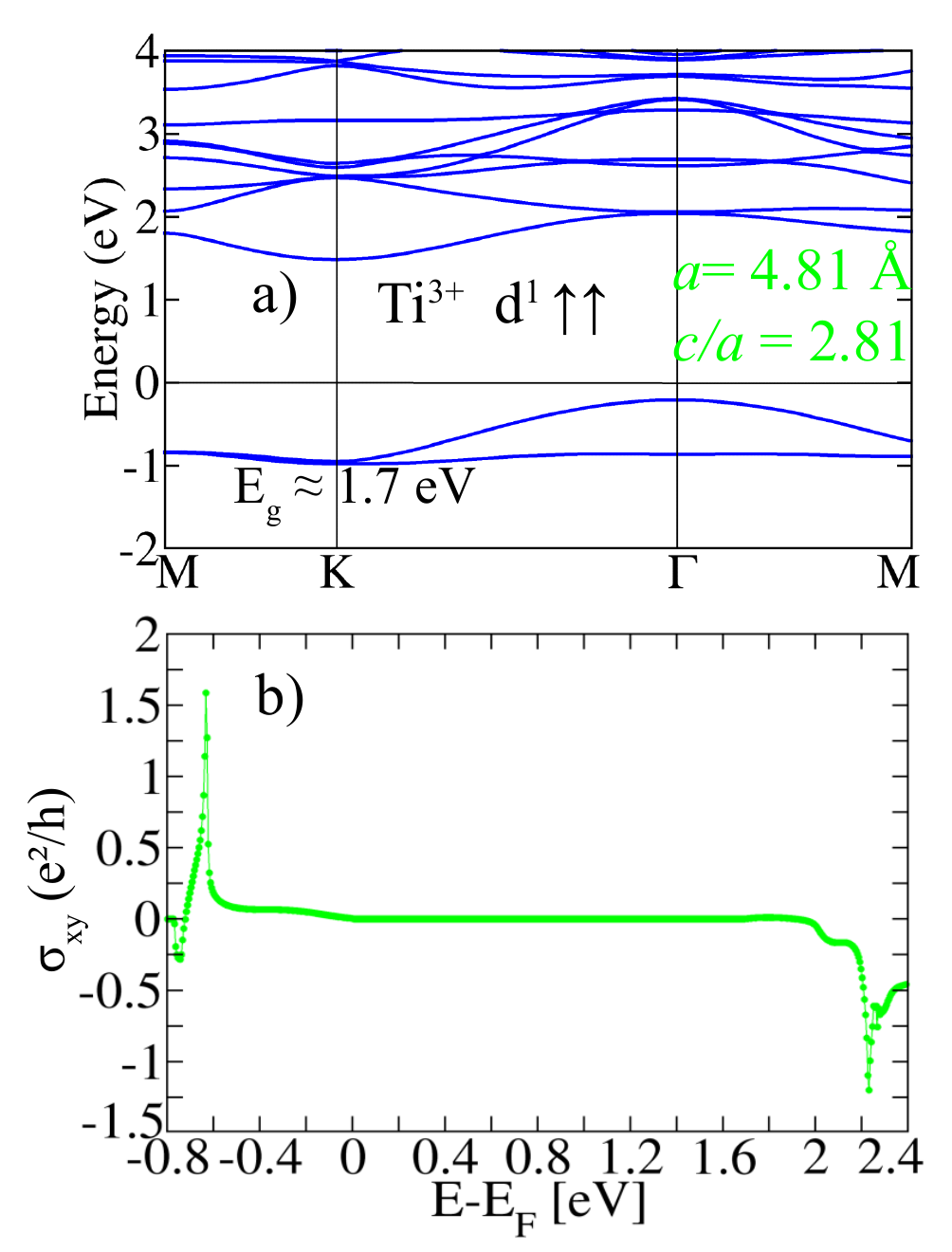}
\caption{Strong effect of SOC for  (Ti$_{2}$O$_{3}$)$_{1}$/(Al$_{2}$O$_{3}$)$_{5}$(0001) at $a_{Al_2O_3}$: a) band structure, b) AHC plotted vs. the chemical potential.}
\label{fig:Ti_Mn_insulating_GS}
\end{figure}

%In a first step we explore the effect of spin-orbit coupling, where we have performed two sets of calculations: one where SOC was included after the converged GGA+$U$ calculations and one where SOC was switched on from the start. The band structures from the first set of calculations are displayed in Fig.~\ref{fig:gga_u_soc}.

Besides the cases presented above, we have also investigated the effect of SOC for $X=$ Ti at the lateral lattice constant of \aalo. In particular, we performed two sets of calculations: one where SOC was included after the converged GGA+$U$ calculations and one where SOC was switched on from the start. The band structure from the first set of calculations is similar to the one displayed in Fig.~\ref{fig:gga_u_soc}a with degenerate solutions for in- and out-of-plane magnetization. In contrast, with the second approach we obtained a particularly strong effect of SOC on the total energy and band structure, although the structural  differences are very small (slight change in the degree of buckling of the honeycomb layer by 0.03\AA). For the Ti$_2$O$_3$-layer the state shown in Fig.~\ref{fig:Ti_Mn_insulating_GS}a with magnetization along (0001) is 0.76 eV per u.c. lower in energy than the one  when SOC is included on top of the converged GGA+$U$ calculation, indicating strong magnetocrystalline anisotropy. The band structure is significantly modified with  the  initial set of four bands now split into two occupied and two empty narrow bands separated by a gap of 1.7 eV. Moreover, a large orbital moment of $-0.88$\mub\ arises, antialigned  and almost compensating  the spin moment of 1.01~\mub.  This case bears analogies to BaFe$_2$(PO$_4$)$_2$ (BFPO)  where a similarly large orbital moment of 0.7~\mub\ was recently reported for Fe$^{2+}$ ($S=2$) \cite{WEP2015}. In this compound Fe$^{2+}$ is also arranged in a honeycomb pattern and has a $d^6$ occupation with completely filled inert spin-up bands and a single electron in the spin-down channel, similar to our case with Ti$^{3+}$ ($3d^1$). 
%The large orbital moment  breaks the degeneracy of the two \egp\ orbitals and enables the opening of a large band gap. 
As shown already in Fig.~\ref{fig:t2g_eg_systems}c) the single electron occupies the degenerate \egp\ orbitals, defined as $\ket{e_{g}'\pm}= \frac{1}{\sqrt{3}}(\ket{d_{xy}}+e^{\pm i\theta}\ket{d_{yz}}+e^{\mp i\theta}\ket{d_{xz}})$, with $\theta=\frac{2\pi}{\sqrt{3}}$, whereas the \ag\ orbital is higher in energy. As discussed recently by Kim and Kee for BFPO\cite{Kee2017}, SOC acts as an atomic orbital Zeeman term: The solution with $L_z=1$ and magnetization along the [0001] direction breaks the degeneracy of the \egp\ orbitals, causing a transfer of charge from the $e_{g}'+$ to the $e_{g}'-$ orbital and thereby opening the significant gap as in BFPO\cite{WEP2015}. Due to the antiparallel alignment of the spin and orbital moment the total moment is almost quenched and (Ti$_{2}$O$_{3}$)$_{1}$/(Al$_{2}$O$_{3}$)$_{5}$(0001) can be regarded as a possible realization of  Haldane's model of  spinless fermions hopping on a honeycomb lattice\cite{Haldane}. The advantage compared to BFPO is that there are no further $d$ electrons, e.g. in the other spin-channel, to interfere with this state. Because the strong SOC effect occurs only for FM coupling, the energy difference between the AFM and FM decreases by an order of magnitude from 814 meV (GGA+$U$) to 56 meV/u.c. (GGA+$U$+SOC), making this intriguing state more likely to be realized. Moreover, a calculation of $\Gamma$-phonon modes indicates an unusual case where strong SOC dynamically stabilizes the system (for more details see Suppl. Material\cite{suppl}). 

A further analogy to BFPO\cite{Kee2017} appears analyzing the band structure and QAH conductivity in Fig.~\ref{fig:Ti_Mn_insulating_GS}a and b: although $\sigma_{xy}$= 0 at \ef\ points to a trivial FM Mott insulator, positive/negative spikes at -0.8 eV and 2.2 eV indicate an occupied/empty set of nontrivial bands with opposite Chern numbers and chirality.  For BFPO Kim and Kee\cite{Kee2017} proposed substitution of Ba$^{2+}$ by mono or trivalent cations in order to shift \ef\ between the lower or upper pair of Chern bands and thereby design a QAHI. This scenario is more difficult to realize for the corundum case, since cations other than trivalent are less common in this structure, but needs to be explored in further studies.
%two copies of nonzero Chern bands, one set above and one set below the Fermi level
 
%For the Mn-bilayer a similarly strong  effect of SOC on the band structure is observed: a gap of 0.65 eV is opened, separating nearly flat bands (cf. Fig. \ref{fig:Ti_Mn_insulating_GS} b). This state is 0.34 eV per u.c. more favorable than the one exhibiting a weak SOC effect. Likewise,  also the energy difference to the AFM ground state is reduced from 0.664 to 0.318 eV. The Mn orbital moment is again antiparallel to the spin moment of $4.09$ \mub, but small ($-0.08$ \mub), still almost an order of magnitude larger than for the case shown in Fig. \ref{fig:gga_u_soc}. Although the anomalous Hall conductivity at the Fermi level is zero  (cf. Fig. \ref{fig:Ti_Mn_insulating_GS}d), the high peaks above and below \ef\ suggest  non-trivial bands.% above t

\section{Summary}\label{sec:fin}

In summary, our DFT+$U$ results on \xoalo(0001) indicate a broad variety of electronic phases w.r.t. orbital polarization and spin state, induced by the confinement of the $3d$ honeycomb layer that are not available in the bulk $X_2$O$_3$ compounds. 
%In particular, a two-dimensional electron gas arises for $X$=Ti and Mn with an impressive degree of confinement, while $X$=Ni shows an insulating state due to disproportionation of the two triangular lattices.
While the ground states in most cases are trivial Mott insulators, followed in stability by symmetry broken ferromagnetic insulating states, for $X$=Ti, Mn, Co and Ni with ferromagnetic coupling and imposed symmetry of the two sublattices, a characteristic  set of four bands with two relatively flat and two crossing at K close to the Fermi level emerges. This feature is similar to the one obtained for $X$= Mn, Co \cite{Doennig2016} and Ni\cite{Yang2011,Ruegg2011,Doennig2014,Doennig2016} in the perovskite \lxolao(111) superlattices  and is a result of the single occupation of the degenerate \egp\ or \eg\ bands. Further analogies arise to the nonmagnetic $d^8$ configuration in $5d$ corundum-derived honeycomb lattices (i.e. $d^4$, single occupation of \eg\ states in each spin channel). Still, the latter topologically nontrivial configuration is metastable (the ground state is found to be AFM for $U>-3$~eV) and transforms into a trivial one with increasing $U$ or tensile strain\cite{Pardo2015}. Further studies are necessary to address the  effect of time-reversal symmetry breaking in $4d$ and $5d$ systems.%Similarly, an AFM ground state emerges also for the perovskite-derived LaAuO$_3$ layer, when correlation effects are included within DFT+DMFT (dynamic mean field theory)\cite{Okamoto2014} or DFT+$U$\cite{Lado2013}. Indeed, also in the $3d$ series of \xoalo(0001) studied the ground state is a trivial antiferromagnetic Mott insulator. For FM alignment, symmetry breaking results also in insulating states. % but with distinct orbital/spin states than the bulk $X_2$O$_3$. %Similarly, insulating orbitally ordered FM phases result, when the imposed symmetry constraint is lifted.   % %crossing at M and not K??

For \xoalo(0001) with $X=$ Mn, Co and Ni and symmetric sublattices, the above mentioned Dirac-crossing  can be tuned to the Fermi level using strain, while for Ti this is hampered by a switching of the orbital polarization from \egp\ to \ag. SOC opens a small gap of several meV in these band structures and leads to a notable anomalous Hall conductivity  at the Fermi level, arising from the sharp peak of the Berry curvature at the avoided band-crossing at K. Overall the tendency towards topological phases is weaker and the gap sizes smaller than in the perovskite bilayers. For comparison, in the LaMnO$_3$ perovskite SL a significant gap of 150 meV is opened by SOC leading to a metastable Chern insulator~\cite{Doennig2016}. 

Finally, for (Ti$_{2}$O$_{3}$)$_{1}$/(Al$_{2}$O$_{3}$)$_{5}$(0001)   we obtain a case of particularly strong SOC at $a_{Al_2O_3}$ that opens a large trivial Mott insulating gap (1.7 eV) with an extremely high orbital moment ($-0.88$\mub) which almost compensates the spin moment $1.01$\mub, thus making the systems a promising candidate for the realization of Haldane's model of spinless fermions on a honeycomb lattice. We note also the presence of nontrivial pairs of bands below and above \ef, suggesting that a Chern insulator may be accessible through suitable electron/hole doping. This strong SOC effect stabilizes dynamically the system and reduces significantly the energy difference to the AFM ground state from 814 meV without SOC to 56 meV.  

The epitaxial growth of corundum films and heterostructures has been less in the focus of investigation compared to perovskites, but offers the advantage that fewer elements are involved (e.g. only $X$ and Al ions, instead of three or four different cations in a perovskite superlattice). 
Several studies have reported the successful growth of corundum thin films on \aalo(0001) using molecular beam epitaxy\cite{Chamberlin2014,Kaspar2014,Dennenwaldt2015}, pulsed laser deposition\cite{Popova2008},  helicon plasma-\cite{Takada2008} or r.f. magnetron sputtering~\cite{Gao2017}. These reports suggest that the growth of corundum-based superlattices is viable and we trust that the  electronic phases predicted here will inspire new experimental studies for their realization.

\begin{acknowledgments}

We acknowledge useful discussions with W. E. Pickett and V. Pardo on honeycomb lattices and Markus E. Gruner on phonon calculations and support by the German Science Foundation (Deutsche Forschungsgemeinschaft, DFG) within the SFB/TRR~80, project G3 and computational time at the Leibniz Rechenzentrum Garching within project pr87ro.

\end{acknowledgments}

\end{document}